\documentclass[prb,twocolumn,floatfix,showpacs,epsf,amssymb]{revtex4-1}
\usepackage[utf8]{inputenc}
\usepackage[english]{babel}
\usepackage{amsmath}
\usepackage{amsfonts}
\usepackage{amssymb}
\usepackage{graphicx}
\usepackage{float}
\usepackage{xcolor}
\usepackage{subfigure}
\usepackage{braket}
\usepackage{mciteplus}

\begin{document}

\title{Multicomponent screening and superfluidity in gapped electron-hole double bilayer graphene with realistic bands}
\author{S. Conti$^{1,2}$, A. Perali$^{3}$, F. M. Peeters$^{2}$, and D. Neilson$^{1,2}$}
\affiliation{$^1$Physics Div., School of Science \& Technology, Universit\`a di Camerino, 62032 Camerino (MC), Italy\\
$^2$Department of Physics, University of Antwerp, Groenenborgerlaan 171, B-2020 Antwerpen, Belgium\\
${^3}$Supernano Laboratory, School of Pharmacy, Universit\`a di Camerino, 62032 Camerino (MC), Italy}
\begin{abstract}
Superfluidity has recently been reported in double electron-hole bilayer graphene. 
The multiband nature of the bilayers is important because of the very small band gaps between conduction and valence bands. 
The long range nature of the superfluid pairing interaction means that screening must be fully taken into account.  
We have carried out a systematic mean-field investigation that includes 
(i) contributions to screening from both intraband and interband excitations, 
(ii) the low-energy band structure of bilayer graphene with its small band gap and flattened Mexican hat-like low-energy bands,  
(iii) the large density of states at the bottom of the bands,
(iv) electron-hole pairing in the multibands, and 
(v) electron-hole pair transfers between the conduction and valence band condensates.
We find that the superfluidity strongly modifies the intraband contributions to the screening, but that the interband contributions are unaffected.  
Unexpectedly, the net effect of the screening is to suppress Josephson-like pair transfers and to confine the superfluid pairing entirely to the conduction band condensate even for very small band gaps, making the system behave similarly to a one-band superfluid.
\end{abstract}

\maketitle

\section{Introduction}

The recent report of enhanced tunneling at equal densities in electron-hole double graphene bilayers \cite{Burg_DBGTunneling_2018} strongly points to the existence of an electron-hole superfluid condensate \cite{Spielman2000}, as predicted theoretically in Ref.\ \onlinecite{Perali_DBG_2013}.  
A number of experimental groups have investigated this system\cite{geim_vanderwaals,lee_giantDrag,li_negativeDrag,Liu2017}.   
It consists of two atomically close, electrically isolated conducting bilayer sheets of graphene, one bilayer with electrons and the other with holes. 
We provide here a systematic theoretical treatment of the competing effects driving and impeding the emergence of a superfluid state, specifically, 
(i) the small band gaps, 
(ii) the non-parabolic shape of the bands with large density of states (DOS), and 
(iii) screening effects from carriers in both the valence and conduction bands.  

The low-energy band structure of bilayer graphene usually has a small band gap that depends in magnitude on the applied perpendicular electric fields \cite{zhang_bandgap} from the metal gates that tune the carrier density \citep{lee_chemical}.     
The shape of the low-energy bands is parabolic for zero gap, but becomes Mexican hat-like when there is a gap: flattened, with a small maximum centered on the $K$ point \cite{Ohta_ARPESgapBG_2006} (see Fig.\ \ref{fig:BG_bands}).
The opening of a gap is accompanied by the development of a large DOS from van Hove-like singularities \cite{Zarenia_Enhancement_2014}.  
The small size of the band gaps, much smaller than the band gaps in conventional semiconductors, suggests that multiband effects cannot be ignored \cite{Conti_MulticomponentDBG_2017}, and in contrast with most studies of superconductors, the long-range Coulomb attraction between the electrons and holes means that screening must be fully accounted for \cite{Lozovik_CorrelationEffects_2012, Sodemann_Interaction_2012, Neilson_Excitonic_2014}.   
References \onlinecite{Perali_DBG_2013,Zarenia_Enhancement_2014,Rios_QMC_2018} included screening but considered only the conduction band of graphene; Ref.\ \onlinecite{Conti_MulticomponentDBG_2017} took into account the valence and conduction bands, but neglected screening. 

The paper is organized as follows.  
In Sec.\ II we recall the physical structure of the electron-hole double bilayer graphene system and its electronic bandstructure.  
In Sec. III we recall multicondensate superfluidity where the superfluid pairs form in more than one band.  
In Sec.\ IV we discuss in some detail, linear screening in a system where there are two graphene bilayers, each with a conduction band and a valence band. 
We discuss the very significant changes in the screening when the system makes a transition from the normal state to the superfluid state.  
In Sec.\ V we present and discuss our results, and Sec.\ VI contains a summary of our main points and our conclusions.
\section{The System}
Our system comprises a pair of electrically isolated graphene bilayers, one bilayer doped with electrons and the other with holes, separated by a thin insulating barrier.  
The dopings can be induced by applying voltages to top and bottom metal gates \cite{Lee_bilayer_doping_2014}.
The bilayers are electrically isolated from each other by insertion of a few atomic layers of insulators like hexagonal Boron Nitride (hBN) \cite{hBN} or WSe$_2$. \cite{Burg_DBGTunneling_2018}
The competing length scales characterizing the system are the barrier thickness, the average interparticle spacing in a bilayer, and the radius of the electron-hole bound pairs.
The energy scales are the Fermi energy, the bandgap between conduction and valence bands, the maximum value of the electron-hole attraction for a given separation of the bilayers, and the magnitudes of the superfluid gaps in the valence and conduction bands.  

The effective Hamiltonian can be written, 
\begin{equation}
\begin{aligned}
H=&\sum_{k,\gamma} \;\left\{\xi^{(e)\gamma}_{k} \,c^{\gamma\dagger}_{k} \, c^{\gamma}_{k} +
\xi^{(h)\gamma}_{k} \,d^{\gamma\dagger}_{k} \, d^{\gamma}_{k} \right\} \\ 
 +& \!\!\!\sum_{\substack{k,k',q\\ \gamma,\gamma'}} \!\!\! \left\{V^D_{k\, k'}
\,c^{\gamma\dagger}_{k+ q/2}
\,d^{\gamma\dagger}_{-k+ q/2}
\,c^{\gamma'}_{k'+ q/2}
\,d^{\gamma'}_{-k'+ q/2} \right.\\ 
 & \hspace{0.4cm} + V^S_{k\, k'}\! \left[
\,c^{\gamma\dagger}_{k+ q/2}
\,c^{\gamma\dagger}_{-k+ q/2}
\,c^{\gamma'}_{k'+ q/2}
\,c^{\gamma'}_{-k'+ q/2} \right.\\ 
& \left.\left. \hspace{1.4cm}\!\!  +\, d^{\gamma\dagger}_{k+ q/2}
\,d^{\gamma\dagger}_{-k+ q/2}
\,d^{\gamma'}_{k'+ q/2}
\,d^{\gamma'}_{-k'+ q/2}\right] \right\}.
\end{aligned}
\label{eq:Hamiltonian}
\end{equation}
We have made the standard transformation for the p-doped bilayer to fill the bands with positively charged holes up to a positive energy Fermi level located in the conduction band.   
The creation and annihilation operators $c^{\gamma\dagger}_{k}$  and $c^{\gamma}_{k}$ are for electrons in the conduction ($\gamma=+$) or valence band  ($\gamma=-$) of the n-doped bilayer.     
$d^{\gamma\dagger}_{k}$ and $d^{\gamma}_{k}$ are the corresponding operators for holes in the p-doped bilayer.  
\begin{figure}[t]
\centering
\includegraphics[width=1\columnwidth]{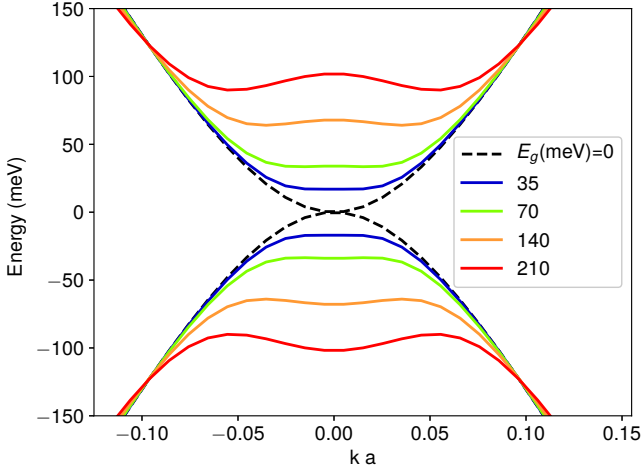} %Fig. 1
\caption{(Color online) Low-energy band structure of bilayer graphene $\varepsilon^{\gamma=\pm}_k$ 
from Eqs.\ (\ref{bandstructure}) and (\ref{bandstructure.supp}), with zero bandgap  (dashed black curve) and with finite bandgap $E_g$ (solid colored curves), as labelled.}
\label{fig:BG_bands}
\end{figure}

$V^S_{k\, k'}$ in Eq.\ (\ref{eq:Hamiltonian}) is the bare repulsive Coulomb interaction between carriers in the same bilayer and $V^D_{k\, k'}$ the bare attractive Coulomb interaction between electrons and holes in the opposite bilayers that are separated by an insulating barrier of thickness $d$, 
\begin{eqnarray}
V^S_{k\, k'} &=& \frac{2\pi e^2}{\epsilon}\frac{1}{|\textbf{k}-\textbf{k}'|}\ ; \nonumber \\
V^D_{k\, k'} &=& -\frac{2\pi e^2}{\epsilon}\frac{e^{-d|\textbf{k}-\textbf{k}'|}}{|\textbf{k}-\textbf{k}'|} \ .
\label{eq:bare_interacions}
\end{eqnarray}
We take the dielectric constant $\epsilon=2$ for bilayer graphene sheets encapsulated in a few layers of insulating hBN. \cite{Kumar_hBN_2016}
 
In Eq.\ (\ref{eq:Hamiltonian}), the energy band dispersions in the tight-binding approximation for a graphene bilayer in AB stacking are given by  \cite{McCann_BilayerGraphene_2013},
\begin{equation}
\varepsilon^{\gamma}_k = (\gamma/2)\left[\sqrt{(t_1-\Gamma_k)^2 +\Omega_k }\right]; \qquad 
\xi^{\gamma}_k = \varepsilon^{\gamma}_k -\mu \ ,
\label{bandstructure}
\end{equation}
where
\begin{eqnarray}
\Gamma_k&=&\sqrt{t^2_1+ 4(\hbar v k)^2 +4E_g^2{(\hbar v k)^2}/{t^2_1}}\ ,\nonumber \\
\Omega_k&=&E_g^2\left[1-4{(\hbar v k)^2}/{t^2_1}\right] \ .
\label{bandstructure.supp}
\end{eqnarray}
Since we are working at low densities for which the carriers occupy only the low energy part of the bands, we retain only two of the four bands (see Fig.\ \ref{fig:BG_bands}).   
We take the bands $\varepsilon^{\gamma}_k$ identical in the two bilayers, and we consider only equal carrier densities, so the chemical potential $\mu$ is the same in both bilayers.  
The tight-binding parameters are $v=\sqrt{3}at_0/ 2\hbar$, intercell distance $a=0.246$ nm, intralayer hopping parameter $t_0\sim 3.16$ eV, and interlayer hopping parameter $t_1\sim0.38$ eV.\cite{Kuzmenko_graphene_2009} 

While the conduction and valence bands of bilayer graphene have certain resemblances to a conventional semiconductor,  they differ in essential respects.   
When a perpendicular electric field is applied across the bilayer by, for example, a potential on a metal gate, a small, variable band gap $0 \leq E_g \lesssim 250$ meV opens between the conduction and valence bands (see Fig.\ \ref{fig:BG_bands}).
The opening of the gap is accompanied by a flattening of the band and development of a small maximum (minimum) in the conduction (valence) band, centered on the $K$ point.  This is the so-called Mexican-hat shape for $\varepsilon^{\gamma}_k$.  

\section{Superfluid state in the multiband system}

Since each bilayer in the electron-hole double bilayer has two bands, pairing in a superfluid can occur between an electron and hole in the conduction band or in the valence band.  
In principle there could also be cross-pairing \cite{Shanenko2015} with the carriers coming from different bands, but we present arguments later that in this system, cross-pairing should not lead to large contributions. 
There are then two main coupled condensates, one  in the conduction band with superfluid gap $\Delta_k^+$, and the other in the valence band with gap $\Delta_k^-$. \cite{Lozovik_Multiband_2009}

At zero temperature within mean-field, the superfluid gaps of the condensates are determined by the coupled equations:
\begin{equation}
\begin{cases}
\Delta^{+}_k \!\!\!\!\! &=-\sum_{k'} \!\! \left[ F^{++}_{kk'} \; V^{e,h}_{k\, k'} \; \frac{\Delta^{+}_{k'}}{2 E^{+}_{k'}} +
F^{+-}_{kk'} \; V^{e,h}_{k\, k'} \; \frac{\Delta^{-}_{k'}}{2 E^{-}_{k'}} \right]\vspace*{2mm} \\
\Delta^{-}_k\!\!\!\!\! &=-\sum_{k'} \!\! \left[  F^{--}_{kk'} \; V^{e,h}_{k\, k'} \; \frac{\Delta^{-}_{k'}}{2 E^{-}_{k'}} +
F^{-+}_{kk'} \; V^{e,h}_{k\, k'} \; \frac{\Delta^{+}_{k'}}{2 E^{+}_{k'}} \right].
\end{cases}
\label{eq:gap}
\end{equation}
where $E^{\gamma}_{k}=\sqrt{(\xi^{\gamma}_{k})^2 + (\Delta^{\gamma}_{k})^2}$, and the geometrical form factor $F^{\gamma\gamma'}_{{kk'}}$ is the overlap of a single-particle state in band $\gamma$ with a state in band $\gamma'$.\cite{Lozovik2010}
We discuss the screened electron-hole interaction between bilayers, $V^{e,h}_{k\, k'}$, below.  

For a given carrier density $n$, the chemical potential $\mu$ is determined from the density equation for the conduction band, 
\begin{equation}
 n= g_s g_v \sum_k (v^+_k)^2 \ ,
\label{eq:density}
\end{equation}
coupled with the gap equations.
We define Bogoliubov amplitudes $v^\gamma_{k}$ and $u^\gamma_{k}$ for the conduction and valence bands,
\begin{equation}
(v^\gamma_{k})^2 = \frac{1}{2}\left(1 -\frac{\xi^{\gamma}_{k}}{E^{\gamma}_{k}} \right)\ ; \quad
(u^\gamma_{k})^2 = \frac{1}{2}\left(1 + \frac{\xi^{\gamma}_{k}}{E^{\gamma}_{k}} \right) \ .
\label{eq:B_a}
\end{equation}

Since we are using the term ``holes'' for the carriers in the p-doped bilayer, to avoid confusion we will refer to an absence of a carrier in the otherwise filled valence band as a ``valence-band vacancy''.  
In the gap equations, Eqs.\ (\ref{eq:gap}), coupling of $\Delta^+_k$ with $\Delta^-_k$ arises from Josephson-like transfers of pairs, where a pair from one band is virtually excited into the other band. 
Pairs that have formed in the valence band can excite into the conduction band, and in the conduction band they reinforce the strength of the $\Delta^+_k$.\cite{Conti_MulticomponentDBG_2017} 
At the same time, these excitations of pairs out of the valence band increases the population of valence-band vacancies.  
The number of valence-band vacancies available to form pairs in the two bilayers controls the strength of the valence band superfluid gap $\Delta^-_k$.  
Since the Fermi energy lies in the conduction band, in the normal state we will start with a negligible population of valence-band vacancies. 
If the Josephson-like transfer is weak, then the superfluid condensates in the two bands decouple, and $\Delta^-_k \ll \Delta^+_k$.  
Thus for weak transfers, the superfluid properties resemble the superfluid in a system with only a conduction band.
In the other limit, when the Josephson-like transfer is strong, the reinforcement of the population of valence-band vacancies by the transfers strongly couples the superfluid condensates, causing $\Delta^-_k$ to approach $\Delta^+_k$ in magnitude.

\section{Screening in a multiband, multilayer system} 
\subsection{Screening in normal state}

The long range nature of the bare Coulomb interaction means that screening of interactions must be taken into account.
With two bilayers, a Coulomb interaction in one bilayer induces a charge response not only in the same bilayer but also in the opposite bilayer.
We use the linear-response random phase approximation (RPA) for screening.  
In the RPA, electrons respond like noninteracting particles to a sum of the external potentials plus the mean-field Hartree potentials from the charge densities induced by the electrons. 
The screened interlayer Coulomb potential in the normal state is \cite{Kharitonov2010}, 
\begin{equation}
V^{eh}_{k\, k'} = \frac{V^D_{k\, k'}}{1 - 2V^S_{k\, k'}\Pi(q)+ \Pi^2(q)\left[(V^S_{k\, k'})^2-(V^D_{k\, k'})^2\right]},\vspace{1mm}
\label{eq:VeffN}
\end{equation}
where $q=|\textbf{k}-\textbf{k}'|$.   
$\Pi(q)$ is the full static RPA polarizability in the multiband bilayer (see Ref.\ \onlinecite{Vignale_book}),
\begin{eqnarray}
\Pi(q) &=& g_s g_v \sum_{\gamma,\gamma'} \Pi^{\gamma\gamma' }(q) \ ,\nonumber \\
\Pi^{\gamma\gamma' }(q) &=& \sum_{k} \frac{\mathit{f}_{k,\gamma}-\mathit{f}_{k',\gamma'}}
{\varepsilon_{k,\gamma}- \varepsilon_{k',\gamma'}}\; F^{\gamma\gamma'}_{{kk'}} \ .
\label{eq:Pi_normal}
\end{eqnarray}
$\mathit{f}_{k,\gamma}$ is the Fermi distribution function for band $\gamma$.   
The bilayer spin and valley degeneracies are $g_s=g_v=2$. 

In Eq.\ (\ref{eq:Pi_normal}), it is useful to distinguish $\Pi^{\mathrm{intra}}(q)$, the intraband contributions in the sum with $\gamma=\gamma'$ for which the stimulus and response are in the same band, 
and  $\Pi^{\mathrm{inter}}(q)$, the interband contributions with $\gamma=-\gamma'$,  for which the stimulus and response occur in opposite bands \cite{Hwang_screening_2008}.  
Refs.\ \onlinecite{DynamicalScreeningBG, BGPolarizability} investigated the separate properties of $\Pi^{intra}(q)$ and $\Pi^{inter}(q)$ for bilayer graphene in the normal Fermi liquid state.  
They showed that at high densities $\Pi^{intra}(q)$ and $\Pi^{inter}(q)$ have qualitatively different dependence on the momentum transfer $q$.  We have now extended this work to low densities. 
Since they were working at high densities, Refs.\ \onlinecite{DynamicalScreeningBG, BGPolarizability}, neglected the small maximum in $\varepsilon^{\gamma}_k$ centered at the $K$ point, the effect of which becomes non-negligible at low densities.

We characterize the different roles played by $\Pi^{intra}(q)$ and $\Pi^{inter}(q)$ in the screening as follows.
For $\Pi^{intra}(q)$, only the conduction band contributes, $\Pi^{intra}(q)\simeq \Pi^{++}(q)$.    
The valence band contribution $\Pi^{--}$(q), is negligible because the valence band is completely full. 
$\Pi^{++}(q)$ scales with the DOS in the conduction band, so $\Pi^{++}(q)=0$ for $n=0$.  
There is a peak in $\Pi^{++}(q)$ at $q=2k_F$, and then for $q>2k_F$ it falls rapidly to zero  \cite{DynamicalScreeningBG}.
This behavior leads to the familiar effect of the screening in real space: 
the screened potential is cut off to zero when $r\agt r_c$, defining a screening length $r_c$. 

In contrast, for $\Pi^{inter}(q)$ the enormous reservoir of carriers in the valence band ensures that $\Pi^{inter}(q)$ is not zero even when the conduction band density $n=0$. 
At $q=0$, we always have $\Pi^{inter}(0)=0$, because interband vertical scatterings and back scatterings are forbidden.
$\Pi^{inter}(q)$ grows monotonically from zero with $q$, and becomes larger than $\Pi^{intra}(q)$ for $q>2k_F$. 
In real space, the large-$q$ behavior of $\Pi^{inter}(q)$ reduces the strength of screened interaction at small $r<r_c$. 
Since $\Pi^{inter}(q)$ involves excitations across the band gap $E_g$, $\Pi^{inter}(q)$ should be sensitive to $E_g$, being strongest for small $E_g$.

\subsection{Screening in superfluid state}

In the presence of superfluidity, the existence of the superfluid gap in the energy spectrum weakens the RPA screened interaction.  
The superfluid condensate reduces the population of free carriers available for screening.  
The RPA screened interaction in the superfluid state is given by \cite{Lozovik_CorrelationEffects_2012,Sodemann_Interaction_2012}, 
\begin{widetext}
\begin{equation}
V^{e,h}_{k\, k'}  = 
\frac{V^D_{k\, k'}  + \Pi_a(q)[(V^S_{k\, k'} )^2-(V^D_{k\, k'} )^2]}
{1- 2[V^S_{k\, k'}  \Pi_n(q) + V^D_{k\, k'}  \Pi_a(q)] + [\Pi_n^2(q) - \Pi_a^2(q)][(V^S_{k\, k'} )^2 - (V^D_{k\, k'} )^2]} \ , 
\label{eq:VeffSF}
\end{equation}
\end{widetext}
where $\Pi_n(q)$ is the normal polarizability that is modified from the polarizability in the Fermi liquid state by the superfluidity, 
\begin{eqnarray}
\!\!\!\!\! \Pi_n(q) &=& \sum_{\gamma,\gamma'} \Pi_n^{\gamma\gamma'}(q) \ ,  \nonumber \\
\label{eq:polarizability}
\!\!\!\!\! \Pi_n^{\gamma\gamma'}\!(q) &=& -\sum_{k} \frac{F^{\gamma\gamma'}_{kk'}}{E^{\gamma}_k+E^{\gamma'}_{k'}} 
\left[ (u^{\gamma}_k v^{\gamma'}_{k'})^2 +(v^{\gamma}_k u^{\gamma'}_{k'})^2 \right],
\end{eqnarray}
and $\Pi_a(q)$ is the anomalous polarizability, \cite{Lozovik_CorrelationEffects_2012,Perali_DBG_2013} which was identically zero in the normal state, 
\begin{eqnarray}
\label{eq:polarizability.a}
\Pi_a(q) &=& \sum_{\gamma,\gamma'} \Pi_a^{\gamma\gamma'}\!(q) \ ,  \nonumber \\
\Pi_a^{\gamma\gamma'}\!(q) &=& \sum_{k} \frac{F^{\gamma\gamma'}_{kk'}}{E^{\gamma}_k+E^{\gamma'}_{k'}} 
(2u^{\gamma}_k v^{\gamma}_{k}v^{\gamma'}_{k'} u^{\gamma'}_{k'})\ .  
\end{eqnarray}
Again $q=|\textbf{k}-\textbf{k}'|$.  
From Eq.\ (\ref{eq:polarizability.a}), we see that $\Pi_a(q)$ has a proportional dependence on the superfluid gaps in the bands, since $\Delta_k^\gamma \propto u^\gamma_k v^\gamma_k$  for band $\gamma$.  
This means that $\Pi_a(q)$ depends on the population of electron-hole pairs in the bands.
For convenience, we again define intra- and interband contributions,
\begin{eqnarray}
\Pi_{n,a}^{intra}(q) &=& \sum_{\gamma}\Pi_{n,a}^{\gamma\gamma}\!(q) \nonumber \\ 
\Pi_{n,a}^{inter}(q) &=& \sum_{\gamma}\Pi_{n,a}^{\gamma,-\gamma}\!(q).
\label{eq:polarizabilityintra&inter}
\end{eqnarray}
\begin{figure*}[!ht]
\centering
\includegraphics[width=1.8\columnwidth]{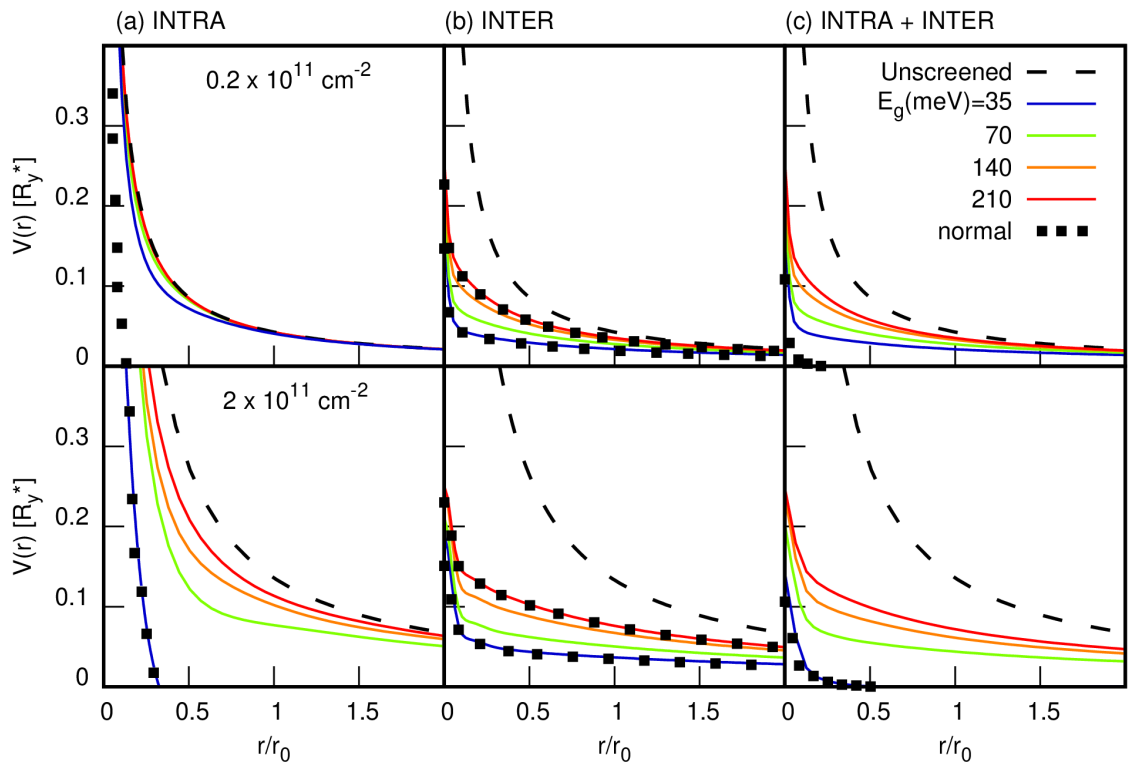} %Fig. 2
\caption{(Color online) Unscreened (dashed lines) and screened electron-hole interaction in real space for the normal state (squares) and the superfluid state (colored curves for different bandgaps $E_g$, as labelled).  
$r_0$ is the interparticle spacing within each bilayer, and Ry$^\star=70$ meV is the effective Rydberg.  
The upper and lower rows show densities $n=0.2\times 10^{11}$ and $n=2\times 10^{11}$ cm$^{-2}$, respectively. 
Column (a) is with only intraband contributions to the screening.
Column (b) is with only interband contributions to the screening.
(For clarity, squares showing screening in the normal state are only given for two values of $E_g$.   
There is the same agreement with screening in the superfluid state for the other colored curves.) 
Column (c) is with both intraband and interband contributions to the screening.}
\label{fig:screening}
\end{figure*}
\subsubsection{One band}

References \onlinecite{Lozovik_CorrelationEffects_2012, Perali_DBG_2013, Neilson_Excitonic_2014} considered only the conduction band, and found within mean-field that superfluidity can significantly weaken screening in a graphene system. 
Recently, quantum Monte Carlo calculations on the system considered in Refs.\ \onlinecite{Perali_DBG_2013, Neilson_Excitonic_2014}, have produced results in good quantitative agreement with the mean-field results \cite{Rios_QMC_2018}. 
Physically, in the superfluid state, the presence of the superfluid gap in the energy spectrum blocks low-lying small-$q$ excitations needed for screening, and superfluid pairing reduces the population of free carriers available for screening. 
Thus in the superfluid state, screening of the long-range interactions is weakened compared with screening in the Fermi liquid state.   
Analytically, within mean-field theory, the reduction in screening is caused by the partial cancellation of the normal and anomalous polarizabilities (Eqs.\ (\ref{eq:polarizability},\ref{eq:polarizability.a})).

In Refs.\ \onlinecite{Perali_DBG_2013, Neilson_Excitonic_2014,Rios_QMC_2018}, no solutions to Eqs.\ (\ref{eq:gap}, \ref{eq:density}, \ref{eq:polarizability}, \ref{eq:polarizability.a}) of physical relevance existed in the weak-coupled BCS superfluid regime with $\Delta \ll E_F$.  
Only in the strong-coupled crossover and BEC regimes, with superfluid gaps $\Delta > E_F$, did solutions exist.  
Physically, this result means that when $\Delta > E_F$, such a wide range of low-lying excited states in the energy spectrum are blocked, that the screening of the electron-hole attractive interaction is sufficiently weakened to allow the superfluidity to exist.  
Further, the large superfluid condensate fraction in the strong-coupled crossover and BEC regimes, means that the population of free carriers available for screening is significantly reduced.
Since the weak-coupled regime would occur at high density, this leads to the prediction of a maximum value of the density for superfluidity to exist, that is, an onset density for superfluidity.  

\subsubsection{Multiband}

Turning to the multiband electron-hole bilayer graphene, Fig.\ \ref{fig:screening} compares the self-consistent screened interaction $V^{e,h}(r)$ between bilayers in real space (see Eq.\ (\ref{eq:VeffSF})) for the superfluid state with the corresponding screened interaction for the normal state.   
The separation of the bilayers is $ d=1$ nm.
Also shown is the unscreened interaction (see Eq.\ (\ref{eq:bare_interacions})).  
$r$ is the component of the electron-hole separation parallel to the bilayers, and $r_0$ is the average interparticle distance within a bilayer.  

Figure \ref{fig:screening}(a) isolates the effect of the intraband screening processes, that is, what the screened interaction $V^{e,h}(r)$ would be if only the $\Pi^{intra}(q)$ contribution to $\Pi(q)$ taken from the full self-consistent calculation, were retained.
At low density, the intraband screened potential in the superfluid state is found to be completely unscreened.
This is because the anomalous polarizability $\Pi_a^{intra}(q)$ fully cancels the normal polarizability $\Pi_n^{intra}(q)$. 
Also shown is the intraband screened potential in the normal state.  
This is completely screened out to zero by $r/r_0\agt 0.3$.
At high density, the cancellation of $\Pi_n^{intra}(q)$ by $\Pi_a^{intra}(q)$ is no longer complete, so the intraband screened potential in the superfluid state is weaker than the unscreened potential.
A new effect in the superfluid state is introduced at this density for the smallest bandgap shown, $E_g=35$ meV: the range of the intraband screened potential in the superfluid state becomes similar to the screened interaction for the normal state.  It is completely cut off by $r/r_0\agt 0.4$.
When the interaction becomes short-ranged, it is found that superfluidity can no longer be sustained. 

Figure \ref{fig:screening}(b) isolates the effect of the interband screening processes, that is, what the screened interaction $V^{e,h}(r)$ would be if only the $\Pi^{inter}(q)$ contribution to $\Pi(q)$ taken from the full self-consistent calculation, were retained.
In contrast to the intraband screening, we see that for interband screening there is no cancellation at all of $\Pi_n^{inter}(q)$ by $\Pi_a^{inter}(q)$.  
Therefore $\Pi^{inter}(q)$ is unchanged from the normal to the superfluid state, and so in the absence of intraband screening, the screened interaction $V^{e,h}(r)$ would be the same in the normal and superfluid states.  
A new effect is that, in contrast with intraband screening, the interband contribution to the screening leads to significant screening in the short-range part of $V^{e,h}(r)$, and this becomes more pronounced with decreasing $E_g$.
The reason for these properties is that the interband contribution to the screening arises from excitations of carriers into the conduction band coming out of the huge reservoir of carriers in the filled valence band, and this becomes stronger as $E_g$ is decreased.  

Finally, Fig.\ \ref{fig:screening}(c) shows the screening in $V^{e,h}(r)$ when both intraband and interband contributions to $\Pi(q)$ taken from the full self-consistent calculation, are included.  We have seen that 
superfluidity cancels out the intraband contributions to the screening, and that it has no effect on the interband contributions.
In the superfluid state at sufficiently high density, the intraband screening contributions to $V^{e,h}(r)$  eventually completely screen out $V^{e,h}(r)$.  
Thus for $E_g=35$ meV, $V^{e,h}(r)$ is completely screened out by $r/r_0 \simeq 0.4$.
The interband contributions weaken $V^{e,h}(r)$ and in this way affect the onset density for superfluidity.  
A smaller $E_g$ results in more interband screening, weakening the electron-hole pairing interactions, and leading to a lower onset density.   
Because of interband screening, when superfluidity does exist, the superfluid gaps are significantly smaller than for the corresponding system with unscreened interactions.
\section{Results}
\subsection{Density dependence of the superfluid gaps}
\begin{figure}[t]
\centering
\includegraphics[trim=2.6cm 0.3cm 3cm 0.5cm, clip=true,width=1\columnwidth]{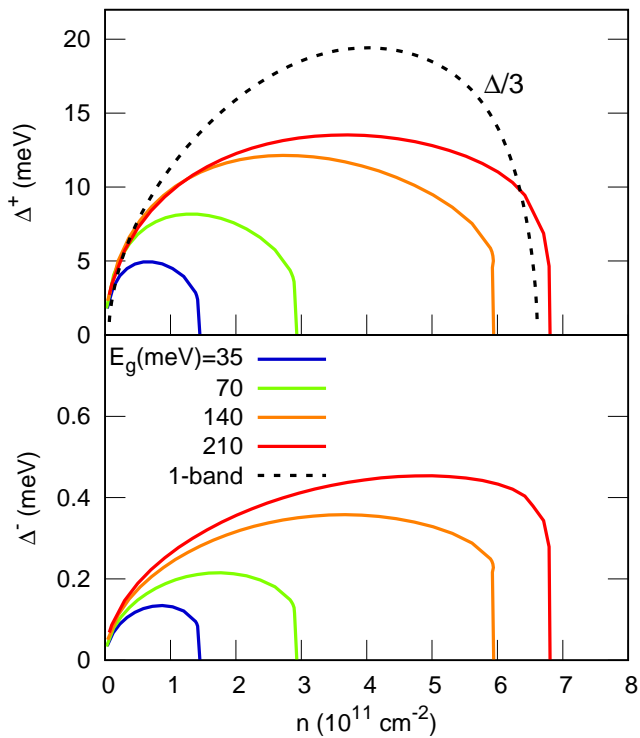} %Fig3
\caption{(Color online) The maximum of the conduction and valence band superfluid gaps $\Delta^\pm=\max_{k}\Delta^\pm_k$ as a function of density for different band gaps $E_g$, as labelled.
Also shown is the maximum of the gap $\Delta=\max_k\Delta_k$ for the corresponding system when only the conduction band is considered, dashed line: $\Delta/3$.}
\label{fig:Delta_max}
\end{figure}
\begin{figure}[t]
\includegraphics[width=1\columnwidth]{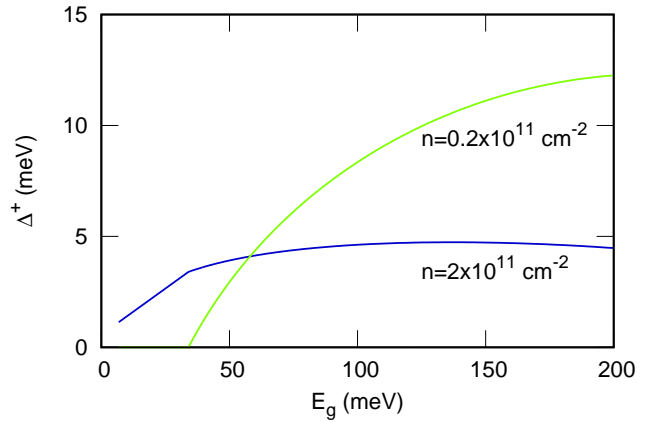} %Fig4
\caption{(Color online) The dependence of the maximum of the superfluid gap in the conduction band $\Delta^+$ as a function of the band gap $E_g$ for two fixed densities.  In all cases, $\Delta^+ \ll E_g$.}
\label{fig:Delta.on.Eg}
\end{figure}

Figure \ref{fig:Delta_max} shows the dependence on density of the maximum of our calculated superfluid gaps $\Delta^\pm=\max_k\Delta^\pm_k$ for the conduction and valence bands (Eqs.\ (\ref{eq:gap})).  
The maxima initially increase with density, since the number of carriers available for pairing is increasing.  
However, at higher densities, the curves pass through a broad maximum and then turn over.
At these densities, screening becomes increasingly effective as the density is increased, so the gaps decrease.   
$E_F$ is increasing with the density, so eventually for the conduction band gap $\Delta^+ < E_F$.  
At around this point, the condensate fraction drops below $\lesssim 0.2$, so there is now a large population of normal-state free carriers available for screening, and the presence of these free carriers enhances the screening. 
Finally, as the density continues to increase, there is a superfluid onset density at which $\Delta^+$ and $\Delta^-$ drop to zero.  
Above the onset density, screening of the electron-hole interactions is so strong that it kills superfluidity.  
For very small band gaps $E_g$, the interband contributions to the screening are strong, and the onset density is very low.
In the other limit, for large band gaps, the onset density is large.

Figure \ref{fig:Delta_max} also shows the maximum of $\Delta$ for the corresponding system with only a parabolic conduction band, as discussed in Ref.\ \onlinecite{Perali_DBG_2013}.
We have taken the effective mass for electrons and holes for the zero gap system, $m^\star_e=m^\star_h=0.04m_e$. \cite{Zou_EffectiveM_2011}
We note in the multiband system, that for larger band gaps $E_g\gtrsim 140$ meV, the predicted density range over which the superfluidity occurs is similar to the density range for the one-band system.
We discuss this result further below.

An unexpected result in Fig.\ \ref{fig:Delta_max} is that, even for large band gaps, $E_g\sim 200$ meV, the conduction band gap $\Delta^+$ remains nearly an order of magnitude weaker than the superfluid gap in the one-band system. 
This is due to the interband contributions to the screening which we have seen are not affected by the superfluidity. 

Another unexpected result in Fig.\ \ref{fig:Delta_max} is that, even for small band gaps $E_g$, the valence band superfluid gap $\Delta^-\ll \Delta^+$ for the conduction band. 
As we have discussed, this result indicates a decoupling of the two gap equations, Eqs.\ (\ref{eq:gap}),  with Josephson-like transfer of pairs always remaining negligible.  
The reason is that the multiband screening always results in superfluid gaps that are much smaller than the band gaps, that is, $\Delta^+ \ll E_g$ (see Fig.\ \ref{fig:Delta.on.Eg}).
It is difficult to generate large $\Delta^+ > E_g$ because the resultant Josephson-like transfer of electron-hole pairs from the valence to the conduction bands would leave in the valence band a significant population of vacancies.  
These free valence-band vacancies would add to the screening and hence reduce $\Delta^+$.   
When the band gap is reduced, the interband screening becomes stronger, which weakens the superfluid gaps.  In this way, the superfluid gap remains smaller than the band gap, $\Delta^+ < E_g$.   

\begin{figure}[t]
\includegraphics[width=1\columnwidth]{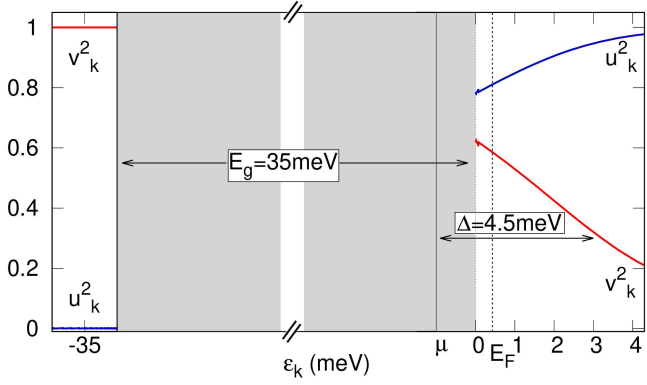}
\caption{(Color online) Bogoliubov amplitudes $u_k$ and $v_k$ as function of energy, for density $n = 1\times 10^{11}$ cm$^{-2}$ and band gap $E_g=35$ meV.}
\label{fig:ukvk}
\end{figure}

To illustrate why Josephson-like transfer of electron-hole pairs are small when $\Delta^{\pm} \ll E_g$, Fig.\ \ref{fig:ukvk} shows the Bogoliubov amplitudes (Eq.\ (\ref{eq:B_a})) for this case. 
The density of valence-band vacancies in the two bilayers available to form pairs in the valence-band condensate, is determined by the overlap of the Bogoliubov amplitudes $v^{-}_k u^{-}_k$.  
Figure \ref{fig:ukvk} shows that this overlap will be vanishingly small whenever $\Delta^{\pm} \ll E_g$, and hence the valence band superfluid gap $\Delta^{-}_k$, which is proportional to $v^{-}_k u^{-}_k$, will be extremely  small.  
If $\Delta^{-}_k$ is small, the cross-coupling term in Eq.\ (\ref{eq:gap}) for $\Delta^{+}_k$ will also be very small.  
Cross-pairing terms, in which superfluid pairs form with carriers from different bands, will also be extremely small because of the vanishingly small population of valence-band vacancies available to contribute to such pairs.  

We can neglect intralayer interactions between carriers within the same bilayer compared with interlayer interactions between electrons and holes in opposite bilayers, for two reasons.
First, attractive interactions between electrons and holes are stronger than the repulsive interactions between like carriers.  Second, the average separation of the electrons and holes, of the order of the barrier thickness, is typically $1$-$3$ nm, while in our density range of interest ($n \lesssim 2\times 10^{11}$ cm$^{-2}$), the average separation of carriers within each layer is much larger, that is, $r_0\gtrsim 13$ nm.

\subsection{Additional effects from the bilayer graphene bands in the presence of a band gap}

\begin{figure}[t!]
\includegraphics[trim=1.3cm 0cm 5cm 0cm, clip=true, width=1\columnwidth]{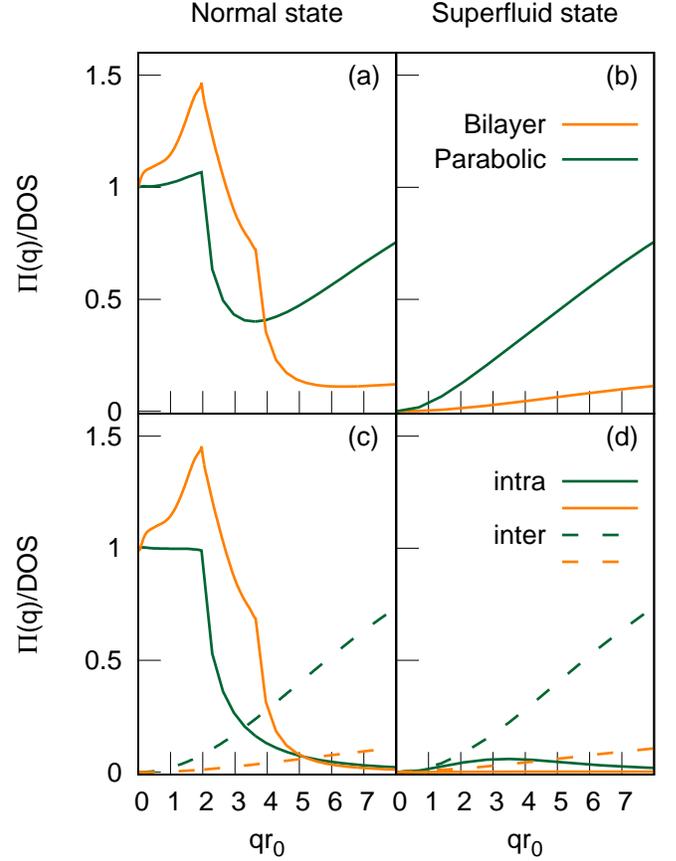} %Fig6
\caption{(Color online) (a) Comparison of polarizability in the normal state using bilayer bands (orange line) and for parabolic bands (green line).  Density $n = 0.25\times 10^{11}$ cm$^{-2}$, band gap $E_g=35$ meV.  
(b) Comparison of the corresponding polarizabilities in the superfluid state. 
(c) Intraband contributions (solid lines) and interband contributions (dashed lines) to the polarizability in the normal state for bilayer bands (orange lines) and for parabolic bands (green lines).
(d) Corresponding intraband and interband contributions to the polarizability in the superfluid state.}
\label{fig:polarizabilty}
\end{figure}
Without a band gap, the conduction and valence bands in bilayer graphene are parabolic at low energies.  
However, the opening of a band gap $E_g$ is accompanied by a flattening of the low-energy  bands and the appearance of a small maximum centered on the $K$ point, the Mexican hat shape \cite{Ohta_ARPESgapBG_2006}.  
The small maximum grows in height with  increasing $E_g$.  
In addition, the DOS around the $K$ point is  strongly enhanced by the development of van Hove-like singularities \cite{Zarenia_Enhancement_2014}. 

The large build-up of the DOS at the bottom of the bilayer conduction band significantly reduces $E_F$ at a given density compared with $E_F$ for the parabolic band, but the flattening of the bands increases $k_F$.  
In addition, at low densities $E_F$ lies below the central maximum of the conduction band, leading to additional effects, discussed below.

Figure \ref{fig:polarizabilty} shows that the polarizabilities $\Pi(q)$ for the normal and superfluid states 
are sensitive to the evolution in the shape of the bands with the development of a band gap.  
Figure \ref{fig:polarizabilty}(a) compares $\Pi(q)$ for the normal state calculated using the bilayer bands  for a small band gap at low density, with $\Pi(q)$ calculated for parabolic bands for the same band gap and density.  
Figure \ref{fig:polarizabilty}(b) makes a similar comparison for $\Pi(q)$ in the superfluid state.  

In the normal state (Fig.\ \ref{fig:polarizabilty}(a)), the polarizability with the bilayer bands is stronger than the polarizability with the parabolic bands over the full range of momentum transfers $q$ that affect screening, $q r_0\lesssim 4$.   
The additional peak in $\Pi(q)$  near $q r_0=2$ for the bilayer bands comes from the small maximum in the conduction band around the $K$ point.  
The peak only appears at densities low enough for $E_F$ to lie below the maximum.  
The maximum generates conduction-band vacancies which add to the intraband screening contribution in this region.  
$\Pi(q)$ then continues larger for the bilayer bands out to $q r_0 \sim 4$, because the flattening of the bands increases $k_F$ for a given density.  

Figure \ref{fig:polarizabilty}(c) separates the intraband and interband contributions to $\Pi(q)$ in the normal state.  
For the bilayer bands, the momentum-transfer range $0\lesssim q r_0\lesssim 5$ is dominated by the intraband contributions, while for $q r_0 > 5$ the interband contributions are larger.
In contrast, for the parabolic bands the intraband contributions dominate only for  $0\lesssim q r_0\lesssim 3$, with the interband contributions larger for $q r_0 > 3$. 
The switch-over from predominantly intraband to predominantly interband screening occurs at a larger $q r_0$ for the bilayer bands because of the flattening of the bilayer bands. 
The flattening increases $k_F$ for a given density compared with the parabolic bands.
The interband polarizability for the bilayer bands is smaller because of their much larger DOS.

In the superfluid state (Fig.\ \ref{fig:polarizabilty}(b)), the polarizability for the bilayer bands is very small for $q r_0 < 4$, while for the parabolic bands it is small only for $q r_0\lesssim 2$. 
The source of this difference is that in the presence of superfluidity, the cancellation between the $\Pi_a(q)$ and  $\Pi_n(q)$ contributions to the screening, only occurs for the intraband screening.
Since  the intraband contribution for the bilayer bands is significant up to $q r_0 \sim 5$ (see Fig.\ \ref{fig:polarizabilty}(c)), the $\Pi_a(q)$ is much more effective in cancelling the screening for the bilayer bands than it is for the parabolic bands, where the screening is suppressed only up to $q r_0 \sim 3$. 
This property also blocks the extra low-lying screening excitations coming from the small maximum at the bottom of the bilayer conduction band that caused the peak near $q r_0=2$ in the normal state $\Pi(q)$ in Fig.\ \ref{fig:polarizabilty}(a).
Once the superfluidity has blocked the intraband screening, what remains is the interband screening.
We have already seen that interband screening is much weaker for the bilayer bands than for the parabolic bands because of the large DOS at the bottom of the bilayer conduction bands.  

To summarize, the primary new effects of the bilayer bands are that 
(i) the intraband contributions dominate out to significantly larger values of $q r_0$ than for parabolic bands, and we recall that only intraband contributions  are suppressed by superfluidity; 
and (ii) the residual interband contributions to the screening are much smaller for bilayer bands than for parabolic bands, because of the large enhancement of the DOS in the low-lying states of the bilayer conduction bands.  

\begin{figure}[ht]
\includegraphics[width=1\columnwidth]{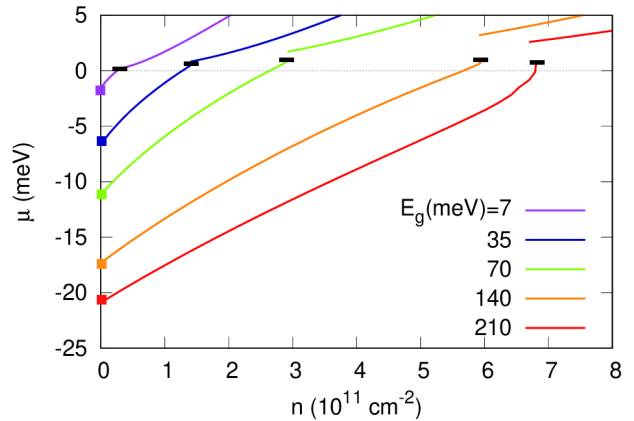} %Fig7
\caption{(Color online) The chemical potential $\mu$ as a function of density $n$. 
The squares mark the low density limiting values $\lim_{n\rightarrow 0} \mu$ (see Table \ref{table:BE}).  The horizontal dashes mark the onset densities at which superfluidity disappears.  
Above the onset density, the system is in the normal state, and thus $\mu=E_F$.}
\label{fig:mu}
\end{figure}

\subsection{One-band superfluidity emerging due to multiband screening}

In the absence of screening, the system naturally divides into two regimes depending on the energy scales \cite{Conti_MulticomponentDBG_2017}: 
(i) for $E_g\gtrsim E_F$, the system resembles a one band system because the contributions from the valence band are negligible;  
(ii) for $E_g\lesssim E_F$, the contribution from the valence band is significant.

However when the electron-hole pairing attraction is screened, the compensatory nature of multiband screening pushes the system to resemble a one-band system, even when the band gap $E_g$ is small.   
We have seen that interband screening keeps $\Delta^+ < E_g$.  
The near complete absence of valence-band vacancies generated by the superfluid, together with negligible Josephson-like pair transfers, keeps $\Delta^-$ very small.   
The large DOS at the bottom of the bilayer conduction band keeps $E_F$ smaller than $E_g$, even for relatively large densities and very small gaps.  

Further independent confirmation of the nearly one-band nature of the superfluidity comes from the behavior of the chemical potential in the limit of small conduction band density, $\lim_{n\rightarrow 0}\mu$ (Fig.\ \ref{fig:mu}).  
For one band, the chemical potential goes to one-half the binding energy of a single electron-hole pair.  
In Ref.\ \onlinecite{Park_TunableExcitons_2010}, the binding energy of an isolated electron-hole pair in a single graphene bilayer, $E_B$, was calculated as a function of band gap $E_g$.   
If the conduction and valence band condensates were strongly coupled, they would become symmetric in the low density limit, in analogy with an intrinsic semiconductor.  
In this case, the limiting value of the chemical potential would lie at the mid-point of the band gap, $E_g/2$, rather than at one-half the binding energy \cite{Conti_MulticomponentDBG_2017}.
In fact, Table \ref{table:BE} shows that $\lim_{n\rightarrow 0}\mu$ in the superfluid state always lies close to the value of $E_B/2$ from Ref.\ \onlinecite{Park_TunableExcitons_2010}, which is behavior consistent with a one-band system.  

\begin{table}
\centering 
\begin{tabular}{|c | c | c | c | c | c|} 
\hline 
$E_g$ &\ 210\ \ &\ 140\ \ &\ 70\ \ &\ 35\ \ \ &\ 7\ \ \ \\ [0.5ex] 
\hline 
$\underset{n\rightarrow 0}\lim\ \mu$ & -21 & -17 & -11 & -6 & -2 \\ 
\hline 
$E_B/2$                               & -23 & -17 & -9 & -5 & -1 \\ [1ex] % [1ex] adds vertical space
\hline 
\end{tabular}
\caption{Comparison of the binding energy $E_B$ of one isolated electron-hole pair in a single graphene bilayer of band gap $E_g$ \cite{Park_TunableExcitons_2010}, with the  low-density limiting behavior of the chemical potential $\mu$ in double bilayer graphene with the same $E_g$, from Fig.\ \ref{fig:mu}. Units are meV.} 
\label{table:BE} 
\end{table}

\section{CONCLUSIONS}

The small band gaps characteristic of bilayer graphene mean that screening by carriers from the filled valence bands strengthens the overall screening.  
This is due to the additional interband contributions to the screening coming from excitations out of the valence band into the conduction band. 

The very large DOS at the bottom of the bilayer conduction band from van Hove-like singularities together with the flattening of the band results, for a given density, in a much smaller Fermi energy $E_F$ than for the parabolic band.
The small Fermi energies permit the superfluidity to be very effective in suppressing screening, with the superfluid energy gap blocking a wide range of low-lying excitations on the scale of $E_F$.  

Despite the small band gaps, Josephson-like pair transfers between the condensates in the valence and conduction bands are negligible.  This unexpected result is because multiband screening always keeps the superfluid gaps small compared with the band gap: any Josephson-like transfer of electron-hole pairs from the valence to the conduction bands leaves behind an increased population of free valence-band vacancies, and these add to the screening.  
The increased screening reduces the superfluid gap.  
The net effect of this compensation is to keep the superfluid gap smaller than the band gap.  

The suppression of  Josephson-like pair transfers means that the superfluid condensates in the valence and conduction bands are decoupled, with the superfluid condensate in the valence band very weak, so that the superfluidity is dominated by the decoupled conduction band condensate.  
The conduction band superfluid gap is significantly weakened by the additional interband screening arising from excitations from the valence band.  
The density range predicted for the superfluidity is consistent with the range predicted in Ref.\ \onlinecite{Perali_DBG_2013}, and of the order of the range reported in recent experiments \cite{Burg_DBGTunneling_2018}. 

Multicomponent screening effects and the evolution of the low-energy bilayer graphene bands with variable band gap, result in a complex interplay of energy and length scales beyond the already rich mean-field results discussed in Ref.\ \onlinecite{Conti_MulticomponentDBG_2017}.  
The comprehensive results presented here, demonstrate the robustness of double bilayer graphene as an optimum platform for realizing and exploiting electron-hole superfluidity under practical experimental conditions.

\subsection*{Acknowledgments} 
This work was partially supported by the Flemish Science Foundation (FWO-Vl) and the Methusalem Foundation.  
We thank Mohammad Zarenia and Alfredo Vargas-Paredes for useful discussions.

\end{document}